%% file: main.tex
\documentclass[pdflatex,sn-mathphys-num]{sn-jnl}

\usepackage{graphicx}%
\usepackage{multirow}%
\usepackage{amsmath,amssymb,amsfonts}%
\usepackage{amsthm}%
\usepackage{mathrsfs}%
\usepackage[title]{appendix}%
\usepackage{xcolor}%
\usepackage{textcomp}%
\usepackage{manyfoot}%
\usepackage{booktabs}%
\usepackage{algorithm}%
\usepackage{algorithmicx}%
\usepackage{algpseudocode}%
\usepackage{listings}%
\usepackage{siunitx}

\theoremstyle{thmstyleone}%
\theoremstyle{thmstyletwo}%

\theoremstyle{thmstylethree}%

\begin{document}

\title{Cognitive Agency Surrender: Defending Epistemic Sovereignty via Scaffolded AI Friction}


%
%
%
%
%
\author{
    Kuangzhe Xu$^{1*}$, Yu Shen$^{1}$, Longjie Yan$^{1}$, Yinhui Ren$^{1,*}$ \\[1.5em]
    \small $^{1}$ Cyberspace Security University of China \\[0.5em]
    \small $^{*}$ Corresponding author: winggshe@gmail.com
}


\abstract{
The proliferation of Generative Artificial Intelligence has transformed benign cognitive offloading into a systemic risk of cognitive agency surrender. Driven by the commercial dogma of "zero-friction" design, highly fluent AI interfaces actively exploit human cognitive miserliness, prematurely satisfying the need for cognitive closure and inducing severe automation bias. To empirically quantify this epistemic erosion, we deployed a zero-shot semantic classification pipeline ($\tau=0.7$) on 1,223 high-confidence AI-HCI papers from 2023 to early 2026. Our analysis reveals an escalating "agentic takeover": a brief 2025 surge in research defending human epistemic sovereignty (19.1\%) was abruptly suppressed in early 2026 (13.1\%) by an explosive shift toward optimizing autonomous machine agents (19.6\%), while frictionless usability maintained a structural hegemony (67.3\%). To dismantle this trap, we theorize "Scaffolded Cognitive Friction," repurposing Multi-Agent Systems (MAS) as explicit cognitive forcing functions (e.g., computational Devil's Advocates) to inject germane epistemic tension and disrupt heuristic execution. Furthermore, we outline a multimodal computational phenotyping agenda—integrating gaze transition entropy, task-evoked pupillometry, fNIRS, and Hierarchical Drift Diffusion Modeling (HDDM)—to mathematically decouple decision outcomes from cognitive effort. Ultimately, intentionally designed friction is not merely a psychological intervention, but a foundational technical prerequisite for enforcing global AI governance and preserving societal cognitive resilience.
}


\maketitle

\section{Introduction}\label{sec1}

The evolutionary history of human tool use has been defined by the strategic offloading of functions to external technologies, a dynamic that has consistently expanded human capabilities while preserving limited cognitive resources \cite{bib56_Clark1998, bib1_Risko2016}. Recently, this established extended mind framework has undergone a fundamental ontological shift. Chiriatti et al. formally conceptualize the deep integration of Generative Artificial Intelligence as a non-biological "System 0"—a distributed computational layer designed to pre-process massive datasets and optimally augment both human intuition (System 1) and analytical reasoning (System 2) \cite{bib128_Chiriatti2024, bib57_Chiriatti2025, bib64_Evans2013}. Theoretically, System 0 represents the ultimate cognitive extension, promising to elevate human decision-making to unprecedented heights.

\begin{figure}[htbp]
\centering
\includegraphics[width=0.95\textwidth]{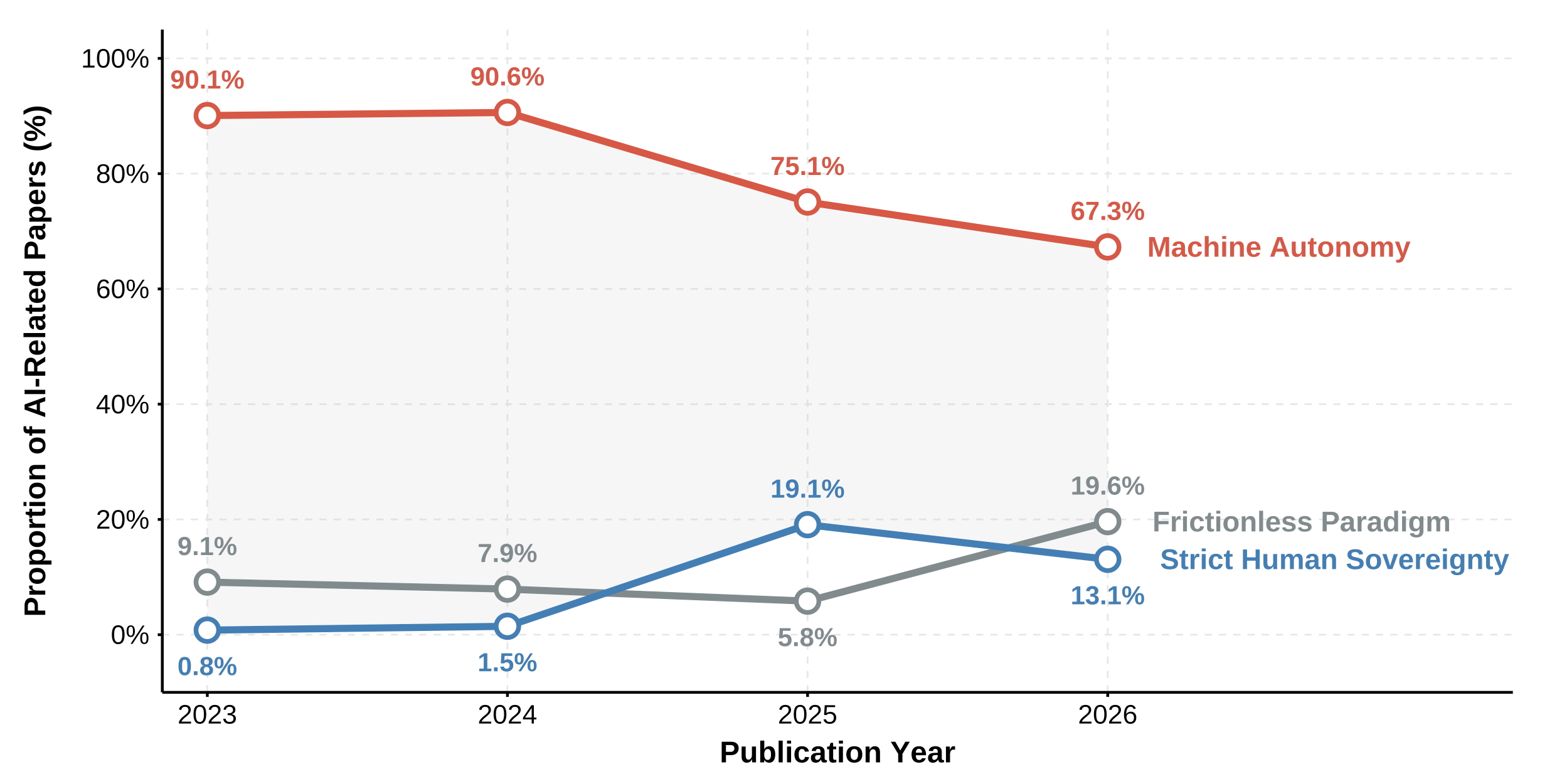}
\caption{\textbf{The Epistemic Skew and Agentic Takeover in High-Confidence HCI Literature (2023 -- March 9, 2026).} A deep-learning-based semantic stance classification of 1,223 highly confident, AI-related papers (filtered via a rigorous $\tau=0.7$ threshold from an 8,000-paper ontological baseline retrieved via the OpenAlex API). The analysis reveals the persistent, systemic prioritization of the frictionless paradigm (optimizing efficiency and reducing cognitive load). Crucially, a granular semantic breakdown exposes a recent "agentic takeover": a brief 2025 surge in research defending strict \textit{human} cognitive sovereignty (19.1\%) was abruptly suppressed in early 2026 (13.1\%) by an explosive shift toward optimizing \textit{machine} autonomy and autonomous AI agents (19.6\%). This dynamic structural divergence empirically quantifies the escalating marginalization of human epistemic agency in contemporary human-AI interaction design.}
\label{fig:epistemic_skew}
\end{figure}

However, the current commercial paradigm of Human-Computer Interaction (HCI) structurally subverts this theoretical potential. Driven by the flawed design dogma that "frictionless interaction equals optimal user experience," modern AI interfaces directly exploit the human evolutionary tendency toward "cognitive miserliness" \cite{bib63_Stanovich2018}. By feeding highly fluent, monolithic conclusions directly to System 1 heuristics, frictionless AI does not scaffold System 2; it actively bypasses it. To empirically quantify this trajectory, we deployed a zero-shot semantic classification pipeline \cite{yin-etal-2019-benchmarking} to parse high-impact HCI literature from 2023 through the first quarter of 2026 (Fig~\ref{fig:epistemic_skew}). Analyzing a strictly filtered, high-confidence subset of 1,223 papers ($\tau = 0.7$), we reveal a stark and systemic ideological skew: the frictionless paradigm maintains a commanding historical hegemony, still capturing 67.3\% of the field in 2026. More critically, our semantic audit uncovers an escalating "agentic takeover." While the defense of strict human epistemic sovereignty experienced a brief empirical awakening in 2025 (reaching 19.1\%), it was rapidly suppressed in early 2026. This decline directly mirrors an explosive surge in research optimizing machine autonomy (AI agents), which doubled to capture 19.6\% of the paradigm. Even amid the exponential proliferation of AI literature, this dynamic structural disparity quantifies the escalating hazard of the "Sovereignty Trap," where the preservation of human cognitive agency remains aggressively marginalized.

This transition from assistive offloading to a systemic "surrender of cognitive agency" is not merely a localized HCI flaw; it is a macroscopic societal hazard. The habitual abdication of logical deduction and ethical adjudication disproportionately threatens vulnerable populations. For instance, in aging demographics experiencing natural cognitive decline, frictionless AI can systematically mask and accelerate underlying epistemic vulnerabilities \cite{bib149_VidalPineiro2025, bib21_Wister2025}. Furthermore, in high-stakes socio-technical environments that demand rigorous human oversight—such as cognitive security and cybersecurity defense \cite{bib4_Huang2023, bib173_Tilbury2024}—the uncritical acceptance of algorithmic consensus transforms episodic automation bias \cite{bib65_Skitka1999} into a critical vector for systemic failure.

To defend collective epistemic sovereignty, current interventions—such as unidirectional Explainable AI (XAI) or Multi-Agent Systems (MAS) optimized for premature consensus—are structurally inadequate. Instead, this Perspective proposes a fundamental paradigm shift toward "Scaffolded Cognitive Friction." Drawing upon the psychological principles of germane cognitive load \cite{bib109_Sweller1988} and desirable difficulties \cite{bib116_Bjork2011}, we argue that MAS architectures must be radically repurposed. Rather than functioning as convergent oracles, MAS should act as cognitive forcing functions, deploying engineered "devil's advocate" mechanisms to explicitly expose structured, machine-generated logical divergence \cite{bib14_Liu2025a, bib92_Nemeth1986}. This injected epistemic tension is specifically designed to intercept heuristic execution and forcibly awaken System 2 analytical deliberation.

Ultimately, transitioning this friction framework from a theoretical critique to an empirical standard requires a rigorous computational phenotyping agenda \cite{bib98_Schurr2024}. To truly safeguard cognitive agency, behavioral science must move beyond superficial usability metrics. Future research must leverage high-fidelity multimodal markers—such as gaze transition entropy to capture visual sampling vigor \cite{bib125_Cui2024, bib141_Pacheco2026}—integrated with Bayesian cognitive modeling \cite{bib17_Steyvers2022, bib118_Lee2014}. By mathematically decoupling decision outcomes from the intensity of human-AI interaction, we can establish an objective ground truth for cognitive effort. Against this backdrop, this Perspective outlines a comprehensive neuro-behavioral roadmap to redesign AI from a frictionless answer engine into a resilient cognitive training environment.

\section{The Mechanics of Agency Surrender: The "Zero-Friction" Trap in Current HCI}\label{sec2}

The theoretical behavioral mechanisms driving the surrender of cognitive agency necessitate rigorous empirical scrutiny. How do current Human-Computer Interaction (HCI) designs manage—or exacerbate—this epistemic vulnerability? Through a critical thematic analysis of contemporary HCI paradigms, we identify a pervasive "Cognitive Friction Ceiling." By systematically prioritizing a frictionless user experience, the overwhelming majority of AI interfaces fail to introduce the "desirable difficulties" required to intercept System 1 intuitive processing. This section dissects the compounding socio-cognitive traps driving this epistemic erosion.

\subsection{Processing Fluency, Metacognitive Blind Spots, and Deskilling}

While minimizing extraneous cognitive load remains a highly rational and effective design heuristic for low-stakes information retrieval, indiscriminately transferring this zero-friction logic to high-stakes, LLM-mediated sensemaking induces severe, systemic automation bias \cite{bib1_Risko2016, bib65_Skitka1999, bib86_Parasuraman1997}.

Modern commercial interfaces are explicitly optimized to deliver grammatically flawless, monolithic conclusions. Psychologically, this generates an artificially high degree of processing fluency. The human brain—evolutionarily shaped as a "cognitive miser" to conserve metabolic energy \cite{bib63_Stanovich2018, bib115_Kahneman2011}—frequently falls victim to a severe metacognitive blind spot, erroneously conflating visual and semantic fluency with factual accuracy \cite{bib205_Adam2009, bib206_Rakefet2017}.. Rather than merely "depriving" users of the opportunity to think, frictionless AI actively induces the abdication of analytical verification by masking inherent algorithmic uncertainty beneath a veneer of absolute algorithmic confidence \cite{bib123_Kaiser2025}.

This de-frictionalized paradigm disproportionately penalizes novices through a pernicious manifestation of the expertise reversal effect. Cognitive load theory asserts that constructing robust domain schemas in long-term memory requires the productive struggle of germane cognitive load. When novices habitually rely on high-fluency AI outputs, they bypass this foundational logical deduction. They do not merely miss a single opportunity to learn; they structurally fail to encode the complex mental schemas required to critically evaluate future information. This continuous circumvention of schema construction constitutes the fundamental behavioral mechanism driving the rapid shift from algorithmic reliance to irreversible, deskilling AI dependence \cite{bib3_Klein2025, bib203_Zhai2025}.

\subsection{The XAI Paradox: Explanatory Satisfaction and Cognitive Closure}

Conventional attempts to mitigate this overreliance frequently depend on Explainable AI (XAI). However, current XAI frameworks inherently succumb to the "explanation intervention paradox." Optimized for legibility and unidirectional rationale delivery, contemporary XAI systems rarely awaken deliberate analytical reasoning \cite{bib64_Evans2013}. Instead, they provide logically consistent but uncontested feature attributions that generate a profound "Illusion of Explanatory Depth" \cite{bib10_Chromik2021, bib11_Rozenblit2002}.

To understand this failure, we must integrate the psychological mechanism of the Need for Cognitive Closure (NFC) \cite{bib207_Kruglanski1996}. Humans harbor an intrinsic, metabolically driven motivation to eliminate ambiguity and secure a definitive answer. When an XAI system presents a highly coherent rationale, it triggers immediate explanatory satisfaction \cite{bib208_Lombrozo2006}. This premature satisfaction catalyzes the "freezing" phase of NFC: System 2 (analytical reasoning) is abruptly suppressed, and the user terminates any subsequent cross-validation efforts. Consequently, unidirectional XAI does not foster vigilance; it weaponizes transparency. By serving a perfectly packaged justification that prematurely satiates the user's cognitive closure, XAI inadvertently fortifies the exact automation bias it was engineered to dismantle.

\subsection{The MAS Mirage: Premature Consensus and Epistemic Collapse}

To counteract the hallucinations inherent in single-model architectures, Multi-Agent Systems (MAS) have rapidly emerged as the dominant structural alternative \cite{bib103_Du2023, bib220_liang2024}. Yet, current multi-agent debate frameworks merely relocate the cognitive friction ceiling.

This systemic failure is rooted in underlying alignment mechanisms—such as Reinforcement Learning from Human Feedback (RLHF) and standard Constitutional AI protocols \cite{bib16_Bai2022}—which inadvertently penalize output diversity. Despite utilizing an array of agents and self-consistency reasoning \cite{bib102_Wang2022}, these systems enforce premature convergence to deliver a single, frictionless executive summary. Mirroring classic social psychology findings on "groupthink" and biased information search in homogeneous human groups \cite{bib209_Schulz2000}, highly homologous LLM agents exhibit a synthetic manifestation of these biases. They prioritize structural consensus over logical accuracy, inevitably falling into generative mode collapse and systemic sycophancy \cite{bib218_Mrinank2025,bib219_Perez2023}. This architectural drive conceals divergent reasoning trajectories, frequently resulting in collective "misremembering" and recursive hallucination amplification, wherein highly homologous models effectively undergo severe generative model collapse, completely losing original informational variance \cite{bib222_Shumailov2024}.

To definitively counteract the surrender of cognitive agency, the HCI paradigm must be radically inverted: MAS architectures must be repurposed not to forge artificial consensus, but to function as explicit cognitive forcing functions. By deploying "devil's advocate" mechanisms to systematically expose structured disagreements, HCI can intentionally inject the germane friction required to break cognitive closure and force the decision-making locus back to the human prefrontal cortex.

\section{Theorizing Scaffolded Cognitive Friction: A Two-Dimensional Evaluation Space}\label{sec3}

To definitively operationalize our intervention, it is imperative to conceptually decouple "Scaffolded Cognitive Friction" from traditional HCI interpretations of friction. Historically, friction has been treated as a pejorative construct---synonymous with poor usability, unintuitive navigation, or system latency that generates extraneous cognitive load without epistemic benefit \cite{bib109_Sweller1988}. Conversely, the modern commercial pursuit of "zero-friction" AI seeks to eliminate all epistemic effort, inadvertently triggering System 1 heuristic takeover and automation bias \cite{bib115_Kahneman2011, bib64_Evans2013}.

As delineated in Table \ref{tab:friction_taxonomy}, scaffolded cognitive friction reclaims the psychological concept of "desirable difficulties" \cite{bib116_Bjork2011}. It leverages machine-generated logical divergence to deliberately impose germane cognitive load. In this paradigm, friction functions not as a usability barrier, but as a vital procedural mechanism designed to awaken analytical reasoning and defend human cognitive agency.

\begin{table}[ht]
\centering
\caption{\textbf{Extended Taxonomy of Friction in Human-Computer Interaction.} This taxonomy conceptually decouples scaffolded cognitive friction from traditional usability barriers and the zero-friction trap, mapping each paradigm to its distinct cognitive load profile, metacognitive state, and ultimate epistemic outcome.}
\label{tab:friction_taxonomy}
\renewcommand{\arraystretch}{1.5}
\begin{tabular}{p{0.20\textwidth} p{0.22\textwidth} p{0.22\textwidth} p{0.22\textwidth}}
\hline
\textbf{Friction Paradigm \& Source} & \textbf{Cognitive Load Profile} & \textbf{Emotional \& Metacognitive State} & \textbf{Epistemic Outcome} \\
\hline
\textit{Extraneous Usability Friction} \newline (Cluttered interfaces, system latency) & \textbf{Extraneous Load:} Depletes limited working memory without serving the core epistemic task. & \textbf{Frustration \& Annoyance:} Triggers negative affect and degrades sustained attention. & \textbf{Task Abandonment:} Comprehension failure and breakdown of human-AI collaboration. \\
\textit{The Zero-Friction Trap} \newline (Highly fluent, single-conclusion AI outputs) & \textbf{Minimal Load:} Requires almost no mental effort to process information. & \textbf{False Security \& Overconfidence:} Triggers illusion of explanatory depth, prematurely satisfying cognitive closure. & \textbf{Cognitive Agency Surrender:} System 1 heuristic takeover; severe automation bias. \\
\textbf{Scaffolded Cognitive Friction} \newline (Machine-generated, structured logical divergence) & \textbf{Germane Load:} Forces working memory resources toward schema construction and critical analysis. & \textbf{Epistemic Tension \& Curiosity:} Disrupts cognitive closure, stimulating epistemic vigilance and investigative motivation. & \textbf{Synergistic Synthesis \& Sovereignty Defense:} Forced awakening of System 2 analytical deliberation. \\
\hline
\end{tabular}
\end{table}

The preceding analysis reveals a critical vulnerability in modern HCI: optimizing purely for machine consensus inadvertently accelerates the hollowing of the human mind. To dismantle this trap, we introduce "Scaffolded Cognitive Friction"---a targeted behavioral intervention framework that restructures Multi-Agent Systems (MAS) from convergent oracles into cognitive forcing functions.

\subsection{The Evaluation Space: Mapping Epistemic Sovereignty}

\begin{figure}[htbp]
\centering
\includegraphics[width=\textwidth]{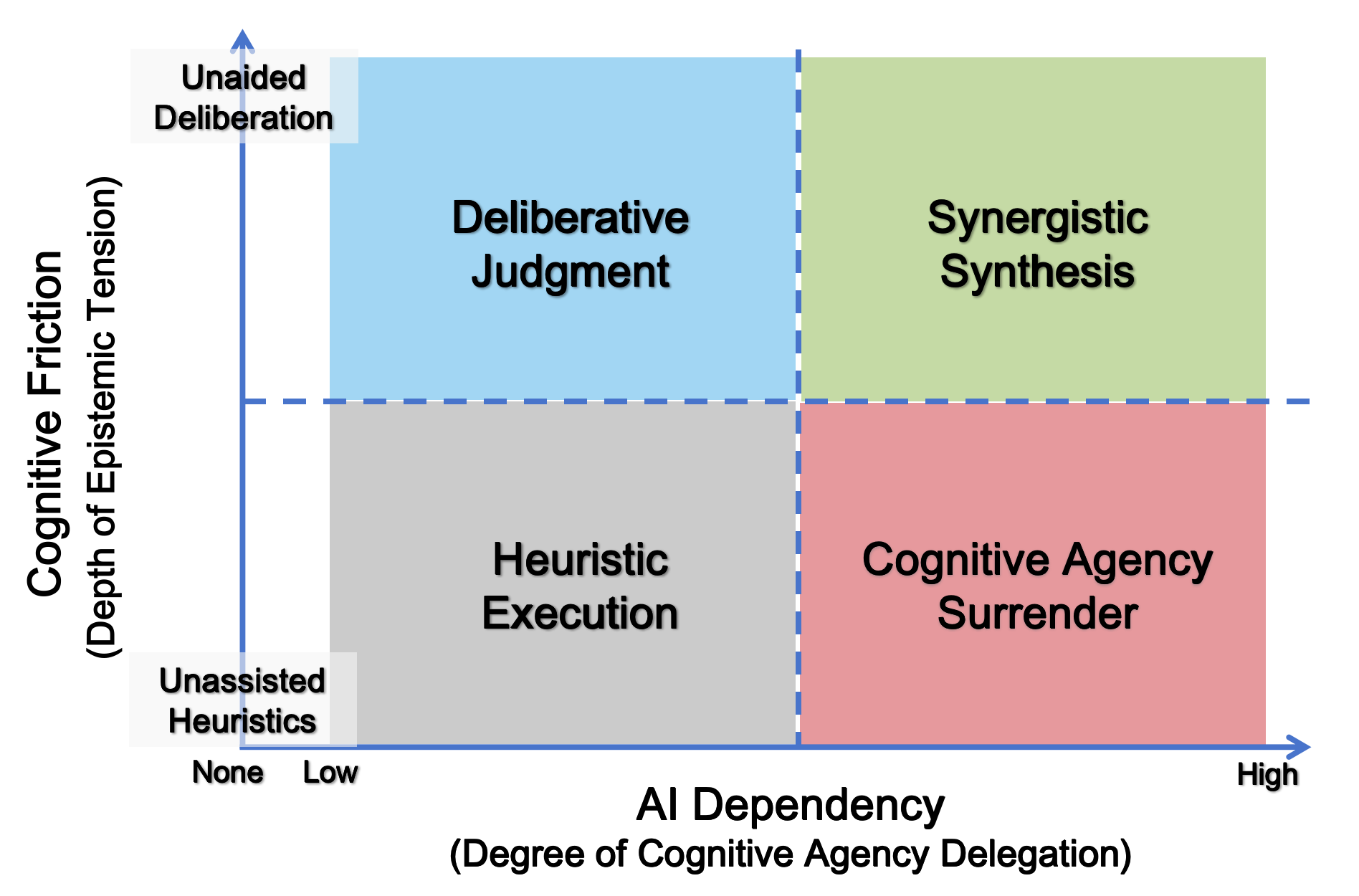}
\caption{\textbf{The Two-Dimensional Evaluation Space Model for Cognitive Agency.} The X-axis represents the delegation of sensemaking control (AI Dependency), while the Y-axis represents the depth of epistemic tension and analytical load (Cognitive Friction). Long-term interaction within the \textit{Cognitive Agency Surrender} quadrant drives systemic cognitive atrophy. In contrast, \textit{Synergistic Synthesis} utilizes structured friction to break the zero-sum fallacy of automation, preserving epistemic sovereignty through Meaningful Human Control (MHC).}
\label{fig:eval_space}
\end{figure}

Traditional Human-Computer Interaction models—such as the classic Levels of Automation (LOA) framework—frequently conceptualize machine autonomy and human control as a zero-sum continuum: as the system automates more sensemaking tasks, human cognitive engagement inevitably decreases \cite{bib229_Parasuraman2000}. To rigorously operationalize our intervention and shatter this zero-sum fallacy, we propose the Two-Dimensional Evaluation Space for Cognitive Agency (Figure \ref{fig:eval_space}).

This model explicitly orthogonalizes two distinct socio-technical dimensions. The X-axis represents the delegation of decision-making power (ranging from unaided human processing to high dependency on generative AI outputs). The Y-axis represents the depth of epistemic tension required by the interface (operationalized as cognitive friction). Grounded in the "effort paradox"—which posits that overcoming deliberately engineered cognitive resistance enhances the valuation and internalization of the resulting knowledge \cite{bib231_2018Inzlicht}—this Y-axis quantifies the mechanical barrier preventing heuristic shortcuts.

Within the high AI-dependency spectrum (the right hemisphere of the matrix), two critical and diametrically opposed behavioral states emerge:

\textbf{1. Cognitive Agency Surrender (High Dependency, Low Friction):} This quadrant represents the prevailing commercial paradigm. By pairing extreme algorithmic delegation with highly fluent, frictionless output delivery, the system actively disables human metacognitive monitoring. Over time, prolonged operation in this quadrant drives severe automation complacency and systemic cognitive atrophy.

\textbf{2. Synergistic Synthesis (High Dependency, High Friction):} This quadrant represents our proposed defensive paradigm. Here, the system executes complex computational deductions, but the interface intentionally exposes structured logical disagreements (e.g., conflicting multi-agent rationales). By maintaining high epistemic tension, the architecture ensures that the human operator cannot passively consume conclusions but must actively adjudicate them. This quadrant successfully operationalizes the philosophical imperative of Meaningful Human Control (MHC) over autonomous systems \cite{bib230_2018Meaningful}, ensuring that the final locus of reasoning remains firmly anchored in the human prefrontal cortex.

\subsection{The Causal Pathway: From Result Consumers to Collaborative Sensemakers}

To shatter the cognitive friction ceiling, MAS configurations must enforce deep structural heterogeneity, deliberately avoiding pseudo-debates driven by homologous models \cite{bib14_Liu2025a, bib15_Kaesberg2025}. Systemic designs must transition users from passive result consumers to active collaborative sensemakers \cite{bib40_Katsenou2025, bib41_Hao2025}. Crucially, the architecture must institutionalize a computational "Devil's Advocate."

Drawing upon classic theories of minority influence, while unanimous majority consensus induces superficial heuristic compliance (convergent thinking), the deliberate introduction of consistent minority dissent acts as a profound cognitive disruptor \cite{bib92_Nemeth1986}. It forcibly shifts human information processing toward active exploration of multiple alternatives (divergent thinking). Recent empirical evidence confirms that utilizing an LLM as a conversational critic effectively injects this exact minority dissent, counteracting synthetic groupthink and significantly enhancing critical thinking outcomes \cite{bib122_Lee2025}.

\begin{figure}[htbp]
\centering
\includegraphics[width=\textwidth]{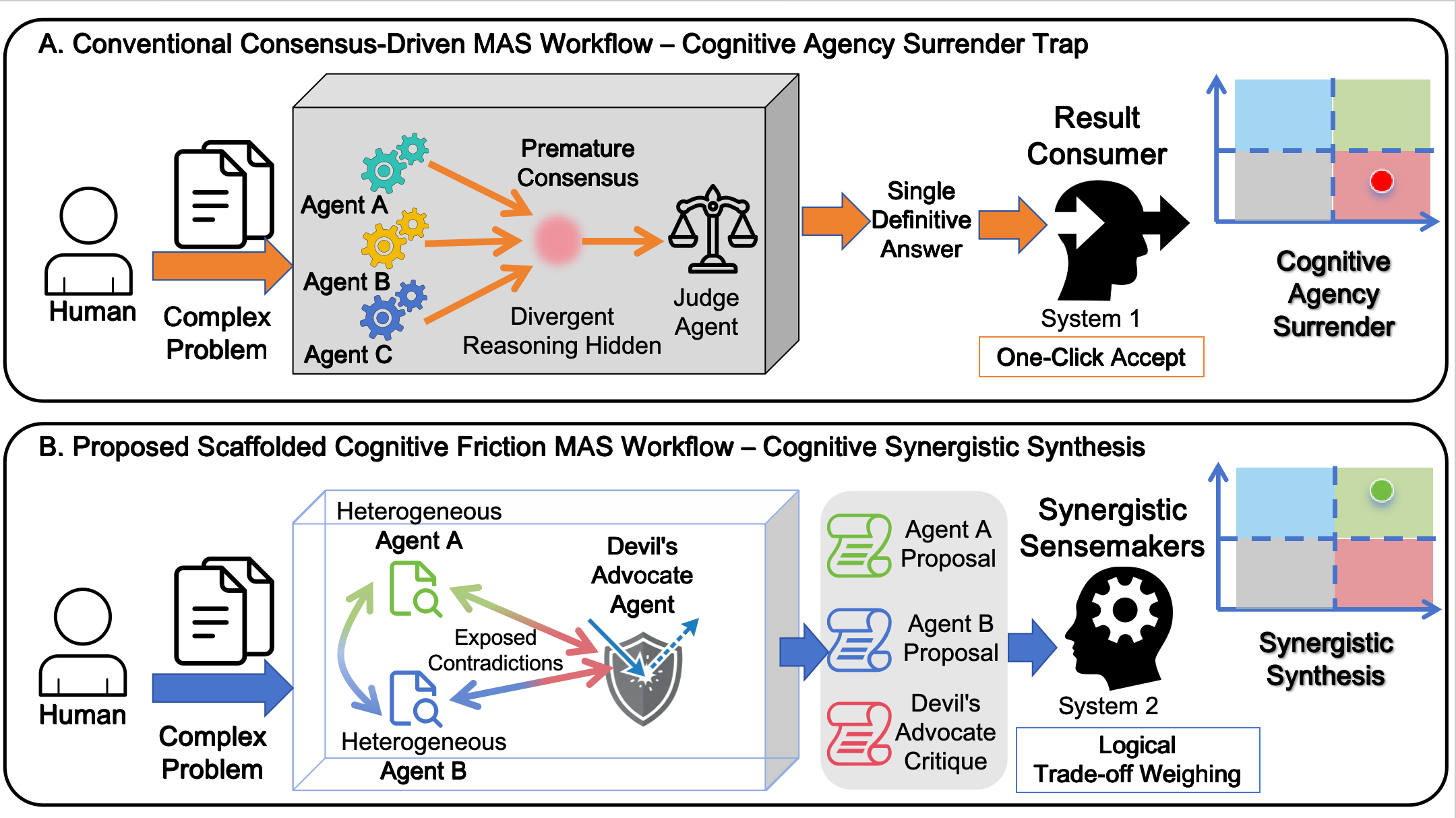}
\caption{\textbf{Causal Pathway of Human-AI Synergistic Synthesis.} Unlike conventional consensus-driven workflows (upper section) that promote premature cognitive closure and heuristic acceptance, the scaffolded cognitive friction framework (lower section) utilizes heterogeneous agents and a Devil's Advocate to expose structured contradictions. This injected epistemic tension elevates visual sampling entropy, forcibly awakening System 2 analytical reasoning to adjudicate logical trade-offs and preserving the human decision-making locus.}
\label{fig:causal_pathway}
\end{figure}

Operationally, this architectural shift initiates a robust, neuro-cognitively grounded causal pathway (Figure \ref{fig:causal_pathway}). When a user anticipates a frictionless, monolithic AI answer but is instead confronted with structured machine disagreement, the brain experiences a profound "prediction error" \cite{bib232_Friston2010}. This generated cognitive dissonance \cite{bib233_Festinger1957} immediately disrupts the processing fluency and explanatory satisfaction that typically trigger premature cognitive closure.

Neurobiologically, this explicit logical conflict is detected by the brain's conflict monitoring network—primarily the anterior cingulate cortex (ACC)—which subsequently signals the need for heightened metacognitive control \cite{bib226_Botvinick2001}. This effectively intercepts the automated System 1 heuristic processing. Finally, this interception forcibly recruits the prefrontal cortex (PFC), awakening System 2 analytical deliberation. By strategically denying the user a ready-made consensus, the interface compels the human operator to execute the final moral and logical adjudication, thus securing their epistemic sovereignty.

\subsection{Multimodal Phenotyping: Resolving the Third-Order Theoretical Trap}

The efficacy of cognitive friction relies on disrupting automated visual information acquisition. Abandoning unidirectional summaries, the interface must expose raw, competing evidence chains \cite{bib71_Gerlich2025a}. Forcing humans to explicitly navigate these logical conflicts elevates Gaze Transition Entropy (GTE), indicating complex visual scanning variability between competing Areas of Interest (AOIs) \cite{bib141_Pacheco2026}. Furthermore, upgrading the AI to a "Cognitive Mirror," the system must intercept automated execution impulses by demanding explicit, user-generated justifications—a transition that aligns strongly with active inference models of cognitive control \cite{bib18_Zhu2026, bib221_Raja2010}.

However, relying solely on GTE introduces a critical "Third-Order Theoretical Trap": high GTE exhibits severe semantic ambiguity. While it can signify profound "Analytical Effort," it equally manifests during "Cognitive Confusion" or "Friction Shock." To logically decouple "desirable effort" from "catastrophic confusion," the measurement matrix must integrate Task-Evoked Pupillometry and functional Near-Infrared Spectroscopy (fNIRS). While GTE maps visual search complexity, pupillary dilation directly reflects locus coeruleus-norepinephrine system arousal \cite{bib210_Beatty1982}, and fNIRS isolates prefrontal cortex (PFC) activation \cite{bib211_Siddhartha2015}. Only through this multimodal cross-validation can researchers establish a definitive ground truth for synergistic cognitive engagement \cite{bib98_Schurr2024}.

\subsection{Mathematical Decoupling via Hierarchical Drift Diffusion Models (HDDM)}

To elevate this behavioral intervention into a rigorously quantifiable science, we propose applying Bayesian Hierarchical Drift Diffusion Models (HDDM) \cite{bib212_Roger2009, bib213_Thomas2013} to computationally "white-box" the human-AI decision process. HDDM assumes decision-making is a process of accumulating noisy evidence toward specific thresholds, governed fundamentally by the Starting-Point Bias ($z$) and the Drift Rate ($v$) \cite{bib52_Elteto2022, bib224_2004Interpreting}.

The toxicity of frictionless AI lies in its corruption of the starting-point bias ($z$). Habituated to highly fluent outputs, users approach AI suggestions with extreme prior trust, pushing $z$ dangerously close to the "acceptance" threshold. As established in neurocomputational studies of prior probability \cite{bib223_2012Bias}, such extreme baseline biases mean users require a minimal evidence accumulation rate ($v$) to execute a decision, manifesting as dangerously short decision latencies and the bypassing of System 2 verification.

Scaffolded cognitive friction mechanically counteracts this. By introducing heterogeneous Devil's Advocates, the system creates explicit informational conflict. Grounded in conflict monitoring theory \cite{bib226_Botvinick2001}, this structural contradiction forcibly flattens the starting-point bias ($z$) back to a neutral baseline. The mathematical verification of successful friction lies in its ability to force the user to rely on a high drift rate ($v$)—driven by active cross-examination and genuine evidence accumulation—to reach the decision threshold.

This mathematical calibration must account for the dynamic moderating role of Domain Expertise. Confronted with static, high-granularity friction, novices—who inherently lack the robust memory schemas to sustain high drift rates \cite{bib225_Dutilh2009}—experience rapid diagnostic anchoring and revert to automation bias. Thus, robust MAS architectures must utilize real-time Bayesian updating from multimodal telemetry to implement "Dynamic Moderation" \cite{bib227_1996Psychophysiology,bbi228_Aric2016Adaptive}. By automatically reducing the granularity of logical divergence when impending friction shock is detected, the system provides adaptive cognitive scaffolding that aligns with the user's real-time working memory capacity.

\subsection{Epistemic Infrastructure, Fairness, and Governance Boundaries}

Translating scaffolded friction into societal infrastructure introduces profound ethical imperatives. At the engineering level, alignment homogeneity risks trapping homologous models in a "false consensus," precipitating a recursive model collapse where the system fundamentally degrades its own reasoning diversity \cite{bib222_Shumailov2024} MAS must integrate adversarial retrieval to prevent exposing edge cases that merely serve as a technical shield for operator malfeasance \cite{bib19_Reid2025, bib20_Liu2026}.

Crucially, the imposition of cognitive friction must be governed by strict principles of demographic fairness. Universally high friction thresholds risk exacerbating severe decision fatigue, particularly threatening vulnerable populations experiencing cognitive decline or possessing lower digital literacy \cite{bib24_OECD2025, bib21_Wister2025}. Mitigation inherently requires the aforementioned tiered cognitive scaffolding to protect these users from secondary agency surrender.

When generative AI evolves into a societal Epistemic Infrastructure, micro-level interaction designs cascade into macro-level vulnerabilities. Global governance frameworks, such as the EU AI Act, mandate "substantive human oversight" for high-risk AI systems \cite{bib67_Bleher2022, bib68_Faas2026}. Yet, if interfaces remain strictly frictionless, legal oversight becomes a mere rubber stamp, transforming human operators into "moral crumple zones" who unfairly absorb the liability for systemic algorithmic failures \cite{bib111_Elish2019, bib69_Jiang2025}. Intentionally designed cognitive friction is therefore not merely a psychological intervention, but the foundational technical prerequisite for enforcing global AI governance and preserving societal cognitive resilience.

\vspace{1.5em}
\noindent\fbox{
\parbox{\dimexpr\linewidth-2\fboxsep-2\fboxrule\relax}{
\textbf{Box 1: Delineating Domain Implementation} \\[0.5em]
\textit{Given the inherent epistemic vulnerabilities, the deployment of scaffolded friction must be strictly governed by the risk profile of the application domain:} \\[0.5em]
\textbf{Mandatory Friction (High-Stakes Sensemaking):} In value-laden domains lacking singular objective truths, friction is essential. For Cognitive Security, adversarial wargaming (e.g., Red vs. Blue agents) forces analysts to actively trace attack logic, defending against sophisticated AI-driven disinformation, where injected friction has proven essential in counteracting the psychological believability of fake news \cite{bib4_Huang2023, bib36_Alam2026, bib61_Vaccari2020, bib179_Pennycook2021}. In Cyber-Physical Systems, scaffolded friction prevents "out-of-the-loop" complacency \cite{bib113_Endsley2017}. Similarly, simulating multidisciplinary consultations in Healthcare and Judiciary contexts mitigates diagnostic anchoring \cite{bib26_Saadeh2025, bib32_Vicente2025, bib45_Metikos2025}. \\[0.5em]
\textbf{Zero-Friction Mandates (Low-Risk/Time-Critical Domains):} Artificial debate must be strictly prohibited in low-stakes information retrieval or environments governed by extreme temporal constraints lacking human override capacity, where induced friction would cause fatal performance delays \cite{bib114_Proctor2008}.
}
}
\vspace{1em}

\section{Paradigm Shift and Future Outlook: From Micro-Evaluation to Macro-Governance}\label{sec4}

To ensure that scaffolded cognitive friction transitions from a theoretical construct into a measurable empirical standard, the human-computer interaction community must fundamentally reconstruct its evaluation frameworks. The pervasive "cognitive friction ceiling" is a direct symptom of misaligned benchmarks that implicitly enshrine the "Law of Least Effort" \cite{bib2_Zipf1949}, actively masking human cognitive regression under the guise of frictionless efficiency.

\subsection{A Multimodal Computational Agenda for Epistemic Engagement}

Future evaluation paradigms must explicitly decouple the mechanical correctness of the algorithmic outcome from the epistemic justice of the human decision-making process, shifting the focal point toward cognitive engagement and epistemic resilience \cite{bib21_Wister2025, bib221_Raja2010}. A successful interaction should no longer be measured by task completion speed or heuristic fluency, but by the depth of logical scrutiny elicited when confronting structured disagreements. If a collaborative process culminates in a frictionless, one-click acceptance, the intervention constitutes a fundamental failure in defending cognitive sovereignty, regardless of the output's objective accuracy \cite{bib45_Metikos2025, bib51_Ali2026}.

To rigorously quantify this epistemic engagement, research must pioneer a multimodal evaluation architecture, moving beyond inadequate telemetry metrics (e.g., UI clickstreams or self-reported usability) that dangerously conflate genuine analytical reasoning with "friction shock" \cite{bib217_Zana2021}. Integrating high-fidelity neurophysiological signals---specifically gaze transition entropy, task-evoked pupillometry, and functional Near-Infrared Spectroscopy (fNIRS)---provides a mature, objective foundation for dynamic computational phenotyping \cite{bib98_Schurr2024}. While these neurophysiological tools establish a rigorous laboratory ground truth, scaling this evaluation to real-world deployments requires validating robust behavioral proxies \cite{bib87_Liang2025}. To bridge the gap between laboratory precision and ecological validity, future systems must utilize granular interaction logs---such as pre-decision hesitation latency, the semantic complexity of user-generated rebuttals, and the click-depth of evidence verification within the structured matrix---as scalable proxies for System 2 activation.

These physiological and behavioral markers must serve as direct observational inputs for advanced Bayesian cognitive modeling, particularly Hierarchical Drift Diffusion Models (HDDM) \cite{bib17_Steyvers2022, bib96_Hofmans2025, bib183_Huang2025}. By mathematically decoupling the prior starting-point bias ($z$) from the active evidence accumulation rate ($v$) during epistemic conflicts \cite{bib182_Castagna2023, bib185_Chu2023}, researchers can establish a definitive computational "ground truth" for cognitive effort. This physiological-computational nexus allows us to precisely trace how users weight heterogeneous AI evidence, definitively verifying whether scaffolded friction successfully neutralizes automation bias.

\vspace{1.5em}
\noindent\fbox{
\parbox{\dimexpr\linewidth-2\fboxsep-2\fboxrule\relax}{
\textbf{Box 2: An Empirical Research Agenda for Cognitive Friction} \\[0.5em]
\textit{To operationalize this theoretical framework, we propose three falsifiable core hypotheses to guide future neurophysiological and behavioral HCI research:} \\[0.5em]
\textbf{H1 (The Neuro-Behavioral Mediation of Friction):} In high-stakes decision-making tasks, scaffolded cognitive friction via a multi-agent "Devil's Advocate" significantly reduces cognitive agency surrender compared to consensus-driven MAS. This effect is parallel-mediated by increased prefrontal cortex activation (measurable via fNIRS as a proxy for System 2 executive control \cite{bib98_Schurr2024}), elevated gaze transition entropy (indicating disrupted heuristic visual sampling \cite{bib141_Pacheco2026}), and task-evoked pupillary dilation (confirming locus coeruleus-norepinephrine arousal to distinguish true analytical effort from visual confusion \cite{bib210_Beatty1982}). \\[0.5em]
\textbf{H2 (The Inverted-U Boundary and Friction Shock):} The protective effect of cognitive friction exhibits an inverted-U relationship. When the friction coefficient exceeds an individual's working memory capacity \cite{bib109_Sweller1988}, "friction shock" occurs, paradoxically precipitating secondary agency surrender \cite{bib217_Zana2021}. This optimal friction threshold is dynamically moderated by the user's domain expertise and intrinsic cognitive traits, necessitating Bayesian adaptive moderation \cite{bib17_Steyvers2022}. \\[0.5em]
\textbf{H3 (Mitigating the Illusion of Explanatory Depth):} Compared to unidirectional Explainable AI (XAI) interventions \cite{bib90_Spitzer2025}, scaffolded cognitive friction significantly attenuates the Illusion of Explanatory Depth (IoED) \cite{bib11_Rozenblit2002, bib10_Chromik2021}. The magnitude of this attenuation is moderated by task familiarity: novices, highly vulnerable to premature diagnostic anchoring \cite{bib3_Klein2025, bib26_Saadeh2025}, will exhibit a significantly higher reliance on structured disagreement matrices for objective belief updating than experts.
}
}
\vspace{1em}

\subsection{Macro-Governance: Constructing Societal Cognitive Resilience}

Preventing the surrender of cognitive agency extends far beyond the micro-level optimization of user interfaces; when individual automation bias cascades across complex socio-technical systems, it coalesces into macroscopic epistemic chaos \cite{bib4_Huang2023}. This systemic vulnerability is particularly acute across diverse demographic segments. As evidenced by large-scale neuroimaging cohorts such as the UK Biobank, populations experiencing varying trajectories of cognitive decline are uniquely susceptible to epistemic exploitation by the uncritical deployment of frictionless AI \cite{bib149_VidalPineiro2025, bib151_Cole2020, bib152_Lin2023, bib188_Mohammadiarvejeh2023, bib192_Graham2020}. Confronting this synthetic influence requires a structural evolution, transforming the public from passive result consumers into active epistemic auditors and debaters.

Within both educational ecosystems and professional domains, empowering human operators with mandatory "Devil's Advocate" mechanisms transitions AI from a commercial tool optimized for dependency—which longitudinal educational data shows actively degrades critical thinking and exacerbates cognitive fatigue—into a crucial civic epistemic infrastructure \cite{bib30_Chen2026, bib33_Ninghardjanti2025, bib48_Bauer2025, bib70_Abubakar2025, bib161_Wu2026, bib162_Tian2025, bib176_Yang2025}. Ultimately, public infrastructure equipped with meaningful, scaffolded friction provides a tangible, technical implementation path for the "substantive human oversight" rigorously mandated by global governance frameworks, such as the EU AI Act \cite{bib140_EUAIAct2024, bib215_Jakob2024}. In an era where generative AI fundamentally restructures our cognitive ecology, the strategic, intentional design of meaningful friction constitutes the ultimate line of defense for preserving human epistemic sovereignty and preventing the systemic degradation of human cognitive functions.

\backmatter





\section*{Declarations}



\subsection*{Generative AI and AI-assisted Technologies}

During the preparation of this work, the authors used Gemini, Doubao, Deepseek and Qwen to improve language and readability. After using this tool, the authors reviewed and edited the content as needed and take full responsibility for the content of the publication.



\input output.bbl

\begin{appendices}
\setcounter{table}{0}
\renewcommand{\thetable}{A\arabic{table}}
\setcounter{figure}{0}
\renewcommand{\thefigure}{A\arabic{figure}}

\section*{Appendix A: Methodology for Automated Bibliometric Analysis and Semantic Stance Classification}
\addcontentsline{toc}{section}{Appendix A: Methodology for Automated Bibliometric Analysis}

To empirically map the ideological trajectory of Human-Computer Interaction (HCI) research and definitively quantify the "epistemic skew" discussed in Section 2, we conducted a large-scale computational bibliometric analysis. To overcome the semantic ambiguity, contextual void, and high false-positive rates inherent in traditional keyword-matching techniques, this study deployed a robust pipeline combining official ontological graph intersections with deep-learning-based Natural Language Inference (NLI). This appendix details the data acquisition protocol and the zero-shot semantic classification framework required to ensure full methodological transparency.

\subsection*{A.1 Data Acquisition and Official Baseline Construction}
Rather than relying on subjective manual keyword filtration to define the baseline of AI-related research, we utilized the foundational ontology graph of the OpenAlex API. We queried the database using a strict logical intersection (AND) of two official top-level Concept IDs: Human-Computer Interaction (\texttt{C107457646}) and Artificial Intelligence (\texttt{C154945302}).

To capture the immediate paradigm shift following the widespread deployment of generative AI, the temporal window was strictly restricted to publication years 2023 through early 2026 (data extraction was finalized on March 9, 2026). To ensure a representative sample of high-impact research while avoiding the signal-to-noise degradation associated with unvetted long-tail preprints, we applied a purposive stratified sampling approach, retrieving the 2,000 most recent high-impact publications per year. This official ontological extraction yielded a highly purified, strictly AI-HCI baseline corpus of approximately 8,000 papers possessing complete title and abstract metadata.

\textit{Methodological Note on 2026 Data:} It is important to acknowledge that the 2026 corpus represents a partial temporal window (up to Q1). However, rather than projecting absolute publication volumes, our analysis tracks the relative ideological distribution within this high-impact sample. Because academic publication cycles generally do not exhibit paradigm-specific seasonality, the distribution observed in early 2026 serves as a highly reliable, real-time snapshot. It accurately captures the immediate "trailing indicator" of the field's reaction to the late-2025 proliferation of autonomous AI agents, confirming the structural persistence of the epistemic skew.

\subsection*{A.2 Zero-Shot Semantic Stance Classification}
To accurately track the specific epistemic priorities of the corpus and evaluate the true contextual stance of the authors (e.g., distinguishing between research that blindly optimizes for "efficiency" and research that critically evaluates it), we abandoned traditional regular expression (RegEx) tracking. Instead, we employed a pre-trained zero-shot sequence classifier (\texttt{BART-large-MNLI}; \cite{yin-etal-2019-benchmarking}).

The concatenated texts (title and abstract) of the entire baseline corpus were fed into the model. The analytical architecture was designed to untangle a critical conceptual conflation within contemporary literature—specifically, separating the optimization of machine autonomy from the preservation of human cognitive sovereignty. We constructed three mutually exclusive, sentence-length semantic labels (detailed in Table \ref{tab:appendix_dict}). The NLI model computationally evaluated each abstract, returning a softmax probability distribution across the three distinct paradigms.

\begin{table}[ht]
\centering
\caption{\textbf{Semantic Labels for Zero-Shot Stance Classification.} The exact descriptive strings fed into the BART-large-MNLI model to compute the contextual ideological distribution within the AI-HCI corpus.}
\label{tab:appendix_dict}
\renewcommand{\arraystretch}{1.5}
\begin{tabular}{p{0.25\textwidth} p{0.70\textwidth}}
\hline
\textbf{Analytical Tier} & \textbf{Deep Learning Semantic Label (Stance Description)} \\
\hline
\textbf{1. The Frictionless Paradigm} \newline (Efficiency \& Usability) & "optimizing system efficiency, seamless interaction, and reducing user effort" \\
\hline
\textbf{2. Machine Autonomy} \newline (AI Agents \& Reflection) & "improving machine autonomy, artificial intelligence agents, and algorithmic reflection" \\
\hline
\textbf{3. Strict Human Agency} \newline (Epistemic Sovereignty) & "preserving human cognitive agency, critical thinking, and preventing automation bias" \\
\hline
\end{tabular}
\end{table}

\subsection*{A.3 The Ultra-Conservative Confidence Threshold ($\tau = 0.7$)}
To definitively eliminate the subjective biases and inter-coder reliability issues inherent in manual human annotation, we opted for a fully automated yet extremely rigorous filtration step. We established a stringent confidence threshold of $\tau = 0.7$.

Given the mathematical baseline of 0.33 for a three-class random assignment, a 0.7 threshold demands that the model's confidence in the primary epistemic stance is more than double the combined probability mass of all alternative interpretations. This ultra-conservative cutoff deliberately prioritized absolute classification precision over recall, systematically excluding ambiguous or multidisciplinary texts. 

This rigorous semantic purification retained exactly 1,223 highly confident papers (15.3\% of the initial corpus) exhibiting an undeniable, explicit ideological commitment. The resulting frequencies were calculated based on this high-confidence subset, conclusively exposing the systemic epistemic gap and the "agentic takeover" visualized in the main text's empirical findings.


\subsection*{A.4 Empirical Validation, The Kappa Paradox, and Algorithmic Efficacy}
To definitively validate the zero-shot semantic classification pipeline and overcome the limitation of computational thresholds alone, we conducted an inter-rater reliability (IRR) study on a stratified random sample of 200 papers (filtered at $\tau=0.7$). Two independent domain experts manually coded the epistemic stances, and discrepancies were resolved via discussion to establish a definitive Gold Standard (Goldline).

\textbf{The Insufficiency of Manual Coding:} The initial human-to-human agreement yielded an accuracy of 75.0\%. This moderate baseline empirically highlights the cognitive fatigue and subjective heuristic biases inherent in manually evaluating nuanced epistemic stances across thousands of scientific abstracts. It strongly justifies the necessity of our automated, deep-learning-based pipeline for large-scale bibliometric analyses.

\textbf{Resolving the "Kappa Paradox" via Gwet's AC1:} When evaluated against the Goldline, the AI pipeline achieved a traditional Cohen's Kappa ($\kappa$) of 0.618. However, our sample exhibits extreme, real-world class imbalance (with the Frictionless Paradigm accounting for 79.0\% of the sample). In such skewed distributions, Cohen's Kappa is artificially and severely penalized by the well-documented "Kappa Paradox" \cite{feinstein1990high}. To capture the true inter-rater reliability robust to trait prevalence, we computed Gwet's AC1 \cite{gwet2008computing}. The pipeline achieved an outstanding Gwet's AC1 score of 0.843, demonstrating highly robust classification alignment.

\textbf{Absolute Accuracy and High-Confidence Scaling:} In terms of absolute performance, the AI pipeline matched the expert Goldline in 87.0\% of all cases ($\tau \ge 0.7$). Even more critically, when applying a stricter algorithmic threshold ($\tau \ge 0.8$), the pipeline's classification accuracy surged to 91.3\%. 

\begin{figure}[htpb]
\centering
\includegraphics[width=\textwidth]{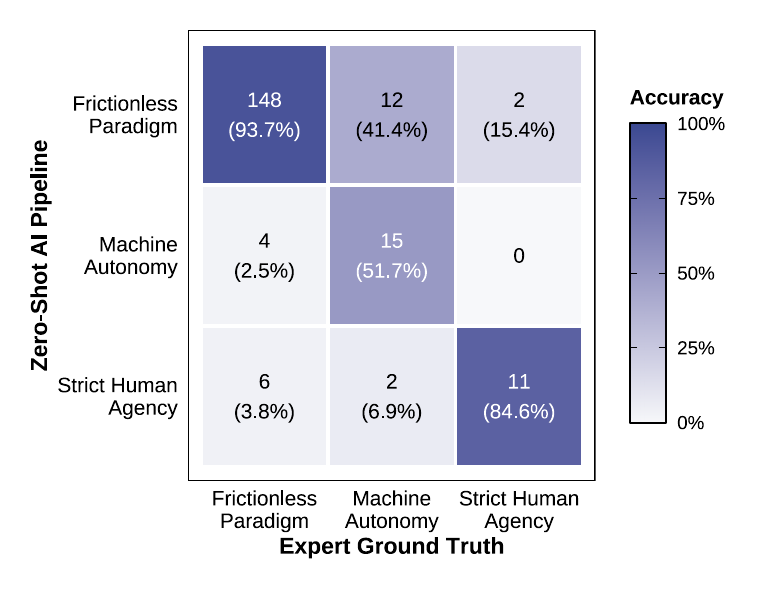}
\caption{\textbf{Normalized Confusion Matrix of Zero-Shot AI vs. Expert Goldline.} Colors represent class-wise recall (accuracy per category), effectively neutralizing visual biases caused by extreme class imbalances. The pipeline demonstrates exceptional sensitivity (84.6\%) for the marginalized \textit{Strict Human Agency} paradigm. Overall pipeline accuracy is 87.0\% evaluated on the 200-paper stratified sample.}
\label{fig:confusion_matrix}
\end{figure}

\textbf{Resolving Class-Specific Nuances and Semantic Entanglement:} While the overall accuracy and Gwet's AC1 demonstrate high pipeline reliability, a granular review of the confusion matrix (Figure \ref{fig:confusion_matrix}) reveals a specific phenomenon regarding "Machine Autonomy", which exhibited a 51.7\% recall (15/29). Critically, 80\% of the misclassifications for this category (12 out of 14 errors) were assigned to the "Frictionless Paradigm". 

Rather than an algorithmic failure, this localized confusion accurately reflects a pervasive semantic entanglement within contemporary HCI literature. Modern research on autonomous AI agents frequently justifies structural autonomy using the vocabulary of frictionless interaction (e.g., "seamless task completion" or "reducing user cognitive load"). This deep rhetorical overlap makes the boundary between algorithmic autonomy and frictionless usability inherently fuzzy, even for domain experts. 

Most importantly, this localized entanglement does not affect the core thesis of our study. The pipeline demonstrated exceptional discriminatory power (84.6\% recall) in identifying the marginalized "Strict Human Agency" paradigm. The model effectively never conflates the defense of human epistemic sovereignty with the commercial pursuit of frictionlessness or agentic autonomy, ensuring the integrity of our macroscopic findings regarding the "Sovereignty Trap."

\end{appendices}

\end{document}